\documentclass[journal,transmag]{IEEEtran}

\usepackage{standalone}
\usepackage[utf8]{inputenc}
\usepackage{etoolbox}
	\newtoggle{arXiv}
	\newtoggle{bibrun}
\usepackage{newtxtext, newtxmath}
\usepackage{graphicx}
\usepackage{color, calc}
\usepackage{array, booktabs}
	\newcolumntype{L}{>{$}l<{$}}
	\newcolumntype{C}{>{$}c<{$}}
	\newcolumntype{R}{>{$}r<{$}}
\usepackage{tikz, tikz-3dplot}
	\usetikzlibrary{arrows, arrows.meta, calc, positioning}
	\usetikzlibrary{decorations.pathmorphing}	
	\tikzset{>=Stealth}
\usepackage{pgfplots}
	\pgfplotsset{compat=1.14, small, scale only axis}
\usepackage{siunitx, physics}
\usepackage{enumitem}
\usepackage{hyperref}

\setlist[description]{labelindent=0pt, leftmargin=\parindent, font=\normalfont\itshape}

\begin{document}
%
\title{Low-field electromagnet for a high-resolution MRI system}


\author{\IEEEauthorblockN{
Juan P. Rigla\IEEEauthorrefmark{1,2},
Franz Bodker\IEEEauthorrefmark{3},
Amir Anari\IEEEauthorrefmark{3},
Eduardo Pallás\IEEEauthorrefmark{2},
Daniel Grau\IEEEauthorrefmark{1,2},\\
Guillermo Puchalt\IEEEauthorrefmark{2},
José M. González\IEEEauthorrefmark{2},
Miguel Corberán\IEEEauthorrefmark{2},
Elena Díaz\IEEEauthorrefmark{1,2},
José M. Algarín\IEEEauthorrefmark{2},\\
Alfonso Ríos\IEEEauthorrefmark{1},
José M. Benlloch\IEEEauthorrefmark{2}, and
Joseba Alonso\IEEEauthorrefmark{2}}

\IEEEauthorblockA{\IEEEauthorrefmark{1}Tesoro Imaging S.L., Ciudad Politécnica de la Innovación, 46022, Spain}
\IEEEauthorblockA{\IEEEauthorrefmark{2}Institute for Instrumentation in Molecular Imaging, Universidad Politécnica de Valencia, 46022, Spain}
\IEEEauthorblockA{\IEEEauthorrefmark{3}Danfysik A/S, Taastrup, 2630, Denmark}

\thanks{Corresponding authors: J. P. Rigla (jpablo.rigla@tesoroimaging.com) and J. Alonso (joseba.alonso@i3m.upv.es).}}

\markboth{Journal of \LaTeX\ Class Files,~Vol.~14, No.~8, August~2015}%
{Shell \MakeLowercase{\textit{et al.}}: Bare Demo of IEEEtran.cls for IEEE Transactions on Magnetics Journals}
%

\IEEEtitleabstractindextext{%
\begin{abstract}
This paper presents the design and experimental characterization of a 1~T electromagnet tailored to meet the demands of a Magnetic Resonance Imaging (MRI) system conceived for spatial resolutions at the level of tens of microns. For high image quality, MRI requires an homogeneous magnetic field over the Field of View (FoV) where the sample is imaged. We measure the relative inhomogeneity of the field generated by our magnet to be well below 100 parts per million over a spherical FoV of 20~mm diameter, while strongly constraining fringe field lines to avoid interference with other devices. The magnet performance closely follows our expectations from numerical simulations in all the experimental tests carried out. Additionally, we present the solutions adopted for thermal management and the design of a mechanical structure to distribute the weight and integrate the platform to move the sample in and out of the magnet.

\end{abstract}

}

\maketitle

%
\IEEEpeerreviewmaketitle

\section{Introduction}

\IEEEPARstart{M}{agnetic} Resonance Imaging (MRI) is a medical imaging modality based on the phenomenon of Nuclear Magnetic Resonance (NMR) \cite{BkBushberg}. As opposed to other imaging techniques, MRI does not require ionizing radiation, but rather exploits a combination of harmless static and Radio-Frequency (RF) magnetic fields. MRI systems generally consist of: i) a magnet to produce a strong, static ``evolution field'' ($B_0$) \cite{13Lvovsky}; ii) an RF system to pulse a time-varying, homogeneous magnetic field ($B_1$) which resonantly excites the spin degree of freedom of nuclei in the sample \cite{07Fujita}; and iii) inhomogeneous magnetic fields which provide the spatial encoding required for imaging \cite{10Hidalgo}. The latter lead to different phase and frequency values for different positions in the Field of View (FoV). Typically, the RF system is designed to address the ubiquitous hydrogen nuclei, forcing a spin precession which in turn induces an ``echo'' signal on the RF coil. This time-dependent signal can be Fourier-Transformed (FT) to obtain images that provide structural and functional information of the sample under study \cite{BkHaacke}.

Image quality in MRI is determined by the interplay of a number of factors, amongst which the evolution-field strength plays a prominent role. The signal strength is largely determined by the so-called magnetization vector ($M_0$), which is given by the abundance of spin-up vs spin-down protons in the sample. At reasonable magnetic-field strengths and sample temperatures, $M_0$ is roughly proportional to $B_0$. In fact, this simplified model fails to capture all the effects that influence the Signal-to-Noise Ratio (SNR), which can grow better than linearly with the field for large values of $B_0$ \cite{98Ocali}. This explains the motivation for building ever more expensive scanners with static fields of $\gtrsim\SI{10}{T}$ \cite{06Vaughan}.

In practice, however, not only is it technically arduous to generate such strong magnetic fields \cite{12Dai}, but it is also challenging to reach the theoretical maximum of SNR in the presence of a strong $B_0$. The use of high magnetic fields also leads to large integrated values of the noise due to the resulting long relaxation times $T_1$ \cite{07Soher}. Moreover, regulators have deferred approval of high-field static systems for safety reasons \cite{14Van}. These include increased RF energy deposition (specific absorption rate, or SAR) at high $B_0$.

In this paper, we present the design and experimental characterization of a low-field electromagnet for a pre-clinical system capable of in-vivo imaging of deep tissue with histological (single cell) resolution (HISTO-MRI Project, \cite{HISTOMRI}). Rather than using the strongest possible evolution field, we combine a highly homogeneous, weak ($\leq\SI{1}{T}$) magnetic field with pulsed, strong magnetic-field gradients of $\sim\SI{10}{\micro s}$ rise and fall times. This will open the door to spatial resolutions as low as \SI{20}{\micro m}, enough to resolve individual cells from one another.

\section{Design of low-field magnet}
\subsection{Main magnet design}
\label{sec:magnet}

For the evolution field $B_0$, the main requirements to meet for the high resolution targeted in the HISTO-MRI project are a field strength from 0.5 to \SI{1}{T} and a spatial homogeneity better than 100~parts-per-million (ppm) in a FoV defined by a spherical region of diameter \SI{20}{mm}. Spatial homogeneity of the field is required to minimize spin-decoherence effects due to different parts of the sample being subject to different magnetic-field strengths and therefore precessing at pseudo-random Larmor frequencies \cite{BkHaacke}. A further, unrelated constraint associated to the laboratory where the experiment sits is that the overall magnet weight must be <\SI{1000}{kg} distributed such as to generate a load <\SI{350}{kg/m^{2}} on the floor (see Appendix). In order to meet these requirements, during the design process we made extensive use of electromagnetics, mechanical and thermal simulations with COMSOL Multiphysics \cite{12Comsol}.

\begin{figure}
	\includegraphics[width= 0.95\columnwidth]{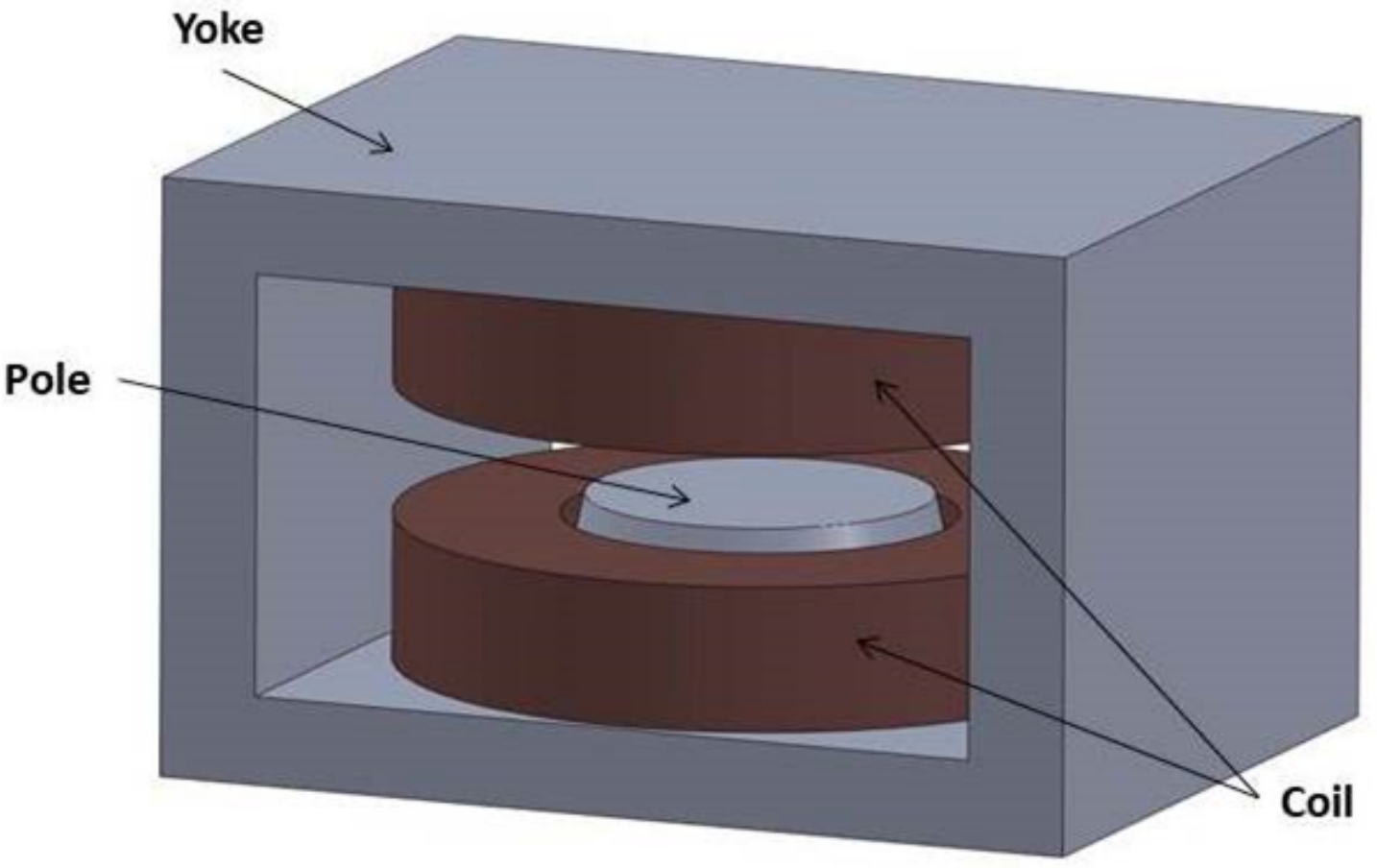}
	\caption{3D magnet model.}
	\label{fig:MagnetModel}
\end{figure} 

We have opted for a classical ``H''-shaped structure (Fig.~\ref{fig:MagnetModel}). This geometry allows to reach higher magnetic fields and offers greater mechanical stability than ``C''-shaped structures. A central part of the design process was to study the effect of pole diameters and gap between poles on the resulting magnetic-field distribution. Without further constraints, the strength and homogeneity requirements were easily met with a magnetic-pole diameter of $\approx\SI{340}{mm}$ and a gap between poles of $\approx\SI{150}{mm}$. However, this led to a total mass >\SI{2700}{kg}, well above what the building can take.

In order to reduce the size and weight of the magnet without sacrificing homogeneity in the FoV, we removed material from the poles and faceted their profile (see Fig.~\ref{fig:MagnetModel}). This is an alternative to the standard solution consisting of a thick iron shim placed at the pole edges (Rose-shim, \cite{18Planche}), which would have reduced the free pole aperture at the edges by around 5~\% and has a poor performance against field excitation. With profile-height variations $<\SI{100}{\micro m}$, the diameter pole was reduced to \SI{250}{mm} and the gap between the magnetic poles to \SI{70}{mm}, resulting in a total mass of \SI{950}{kg}, which we redistribute over enough surface with a dedicated mechanical structure (see Appendix).

\begin{figure}
	\includegraphics[width= 1\columnwidth]{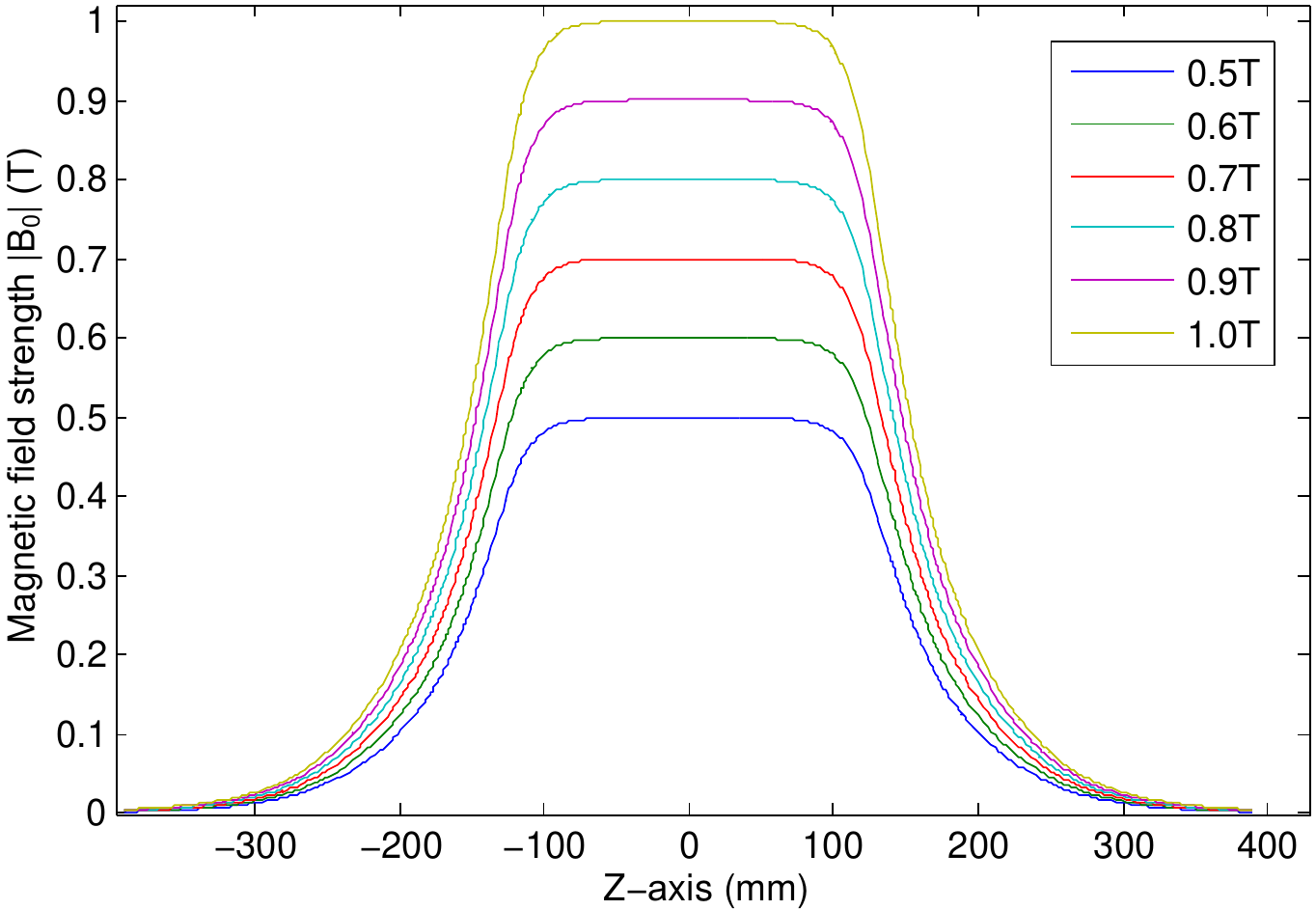}
	\caption{Z-axis distribution of the magnetic-field magnitude $\abs{B_0(z)}$ for maximal field strengths ranging from 0.5 to \SI{1}{T}.}
	\label{fig:MagnetFieldStrengths}
\end{figure} 

In order to generate the field, the magnet counts with two circular coils of 156 turns each, with a total conductor cross section of \SI{51}{mm^2}. With this design, a current of $\approx\SI{180}{A}$ through the coils results in the desired maximal field strength of \SI{1}{T} (yellow curve in Fig.~\ref{fig:MagnetFieldStrengths}).

\begin{figure}
	\includegraphics[width= 1\columnwidth]{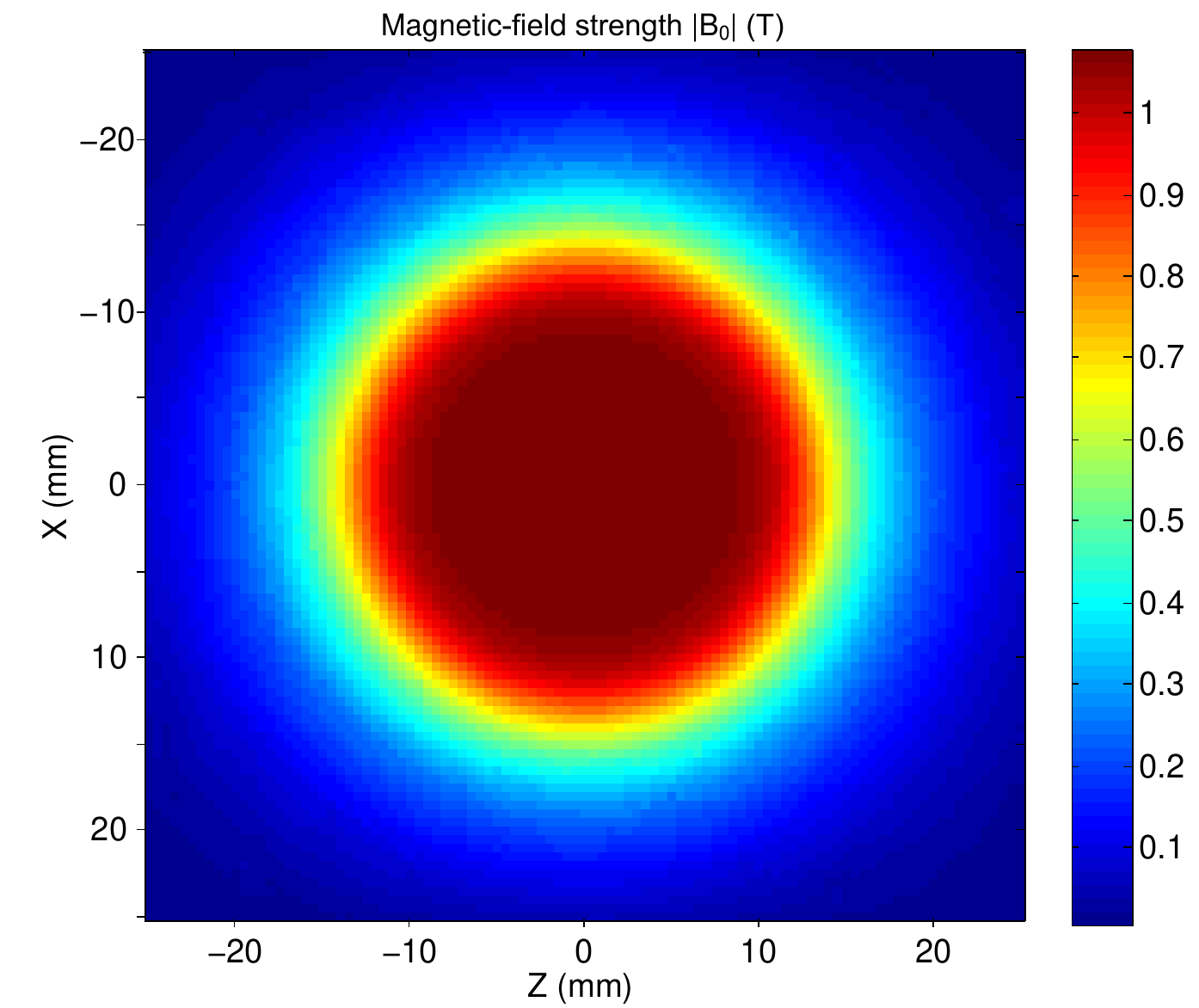}
	\caption{Magnetic-field strength (in Tesla) in the $Y=0$ plane for a maximal field strength of \SI{1}{T}.}
	\label{fig:MagnetDistXZ}
\end{figure}

The magnetic field profiles in the median ($Y=0$) plane for a field strength of \SI{1}{T} are shown in Fig.~\ref{fig:MagnetDistXZ}. Numerical simulations indicate a Root-Mean-Square (RMS) homogeneity of \SI{68}{ppm}, based on the deviation of the field strength from that at the center of the FoV. In order to quantify the influence of possible manufacturing and assembly errors, we simulated an intentional pole-gap deviation of \SI{20}{\micro m} and found an RMS homogeneity of \SI{286}{ppm}.

\subsection{Magnetic-shielding design}

\begin{figure*} 
	\centering
	\includegraphics[width= 1.95\columnwidth]{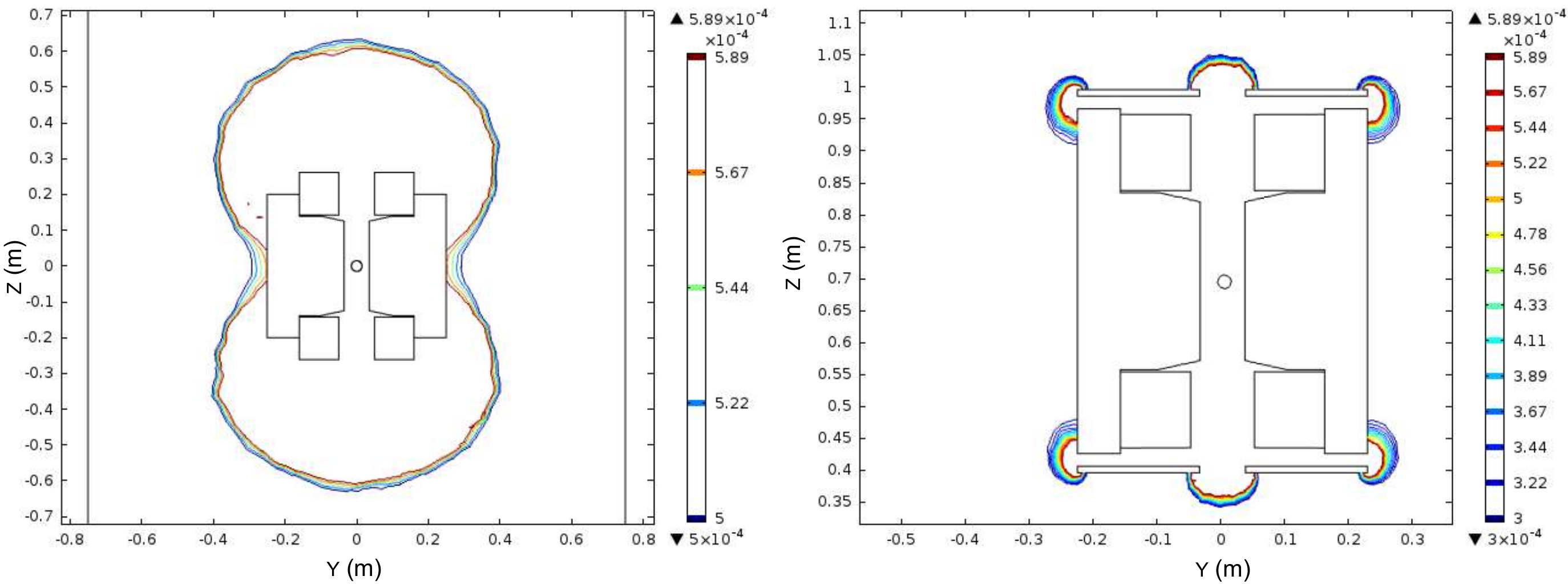}
	\includegraphics[width= 1.95\columnwidth]{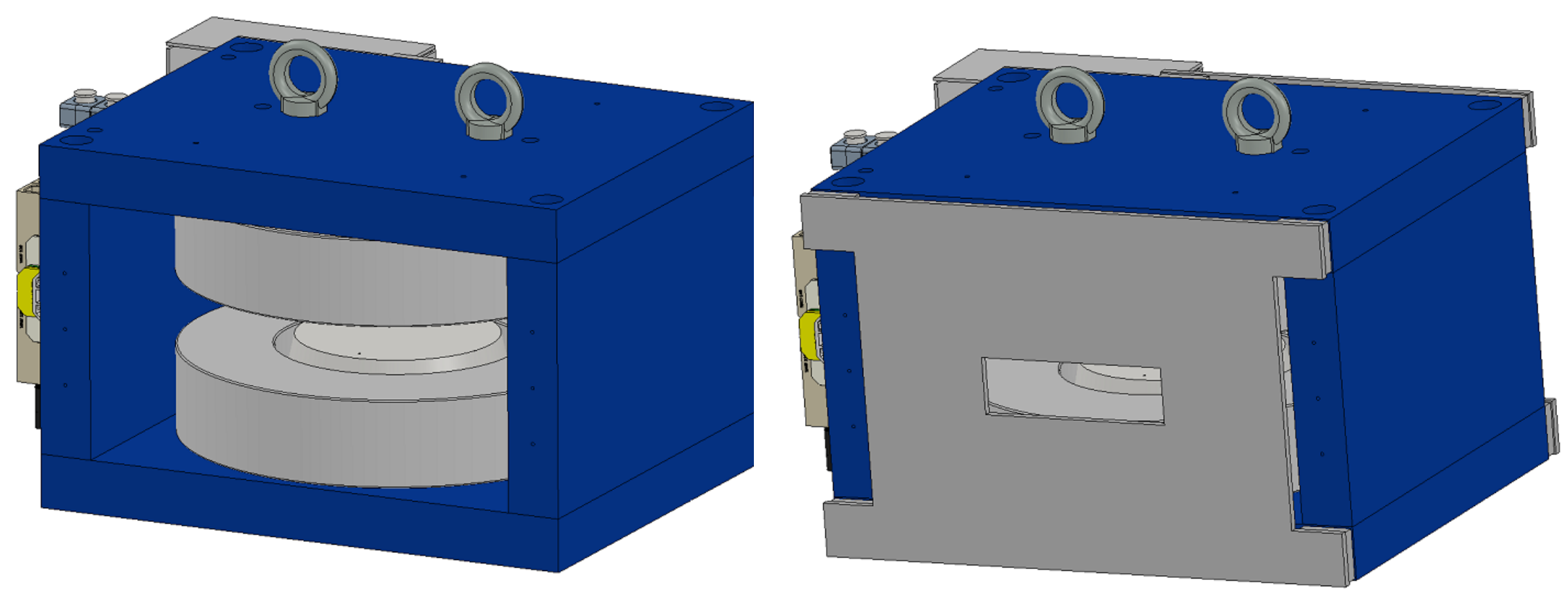}
	\caption{(Top) Fringe-field distribution (in Tesla) in the YZ plane without (left) and with (right) shielding steel-plates. Note the different length scales in the plots. The circles indicate the center of the FoV. (Bottom) 3D models of the magnet  without (left) and with (right) shielding steel-plates.}
	\label{fig:FringeDist}
\end{figure*}

The presence of the magnet will result in unwanted and potentially dangerous effects on other components and laboratory equipment if the peripheral fringe-field outside the core is sizable. Fringe fields are typically characterized by the extent of the 5~Gauss surface. For the design in Fig.~\ref{fig:FringeDist} (top left) this reaches \SI{610}{mm} from the front and back of the bare magnet if operated at \SI{1}{T}, well beyond desired.

In order to address this issue, we considered placing the magnet inside a high-permeability box available to us. This consists of two steel layers of thickness \SI{1}{mm} and separated by a wooden layer of thickness \SI{20}{mm}. Simulations indicated the magnetic field outside the box would be $<\SI{2}{G}$. Nevertheless, we abandoned this approach due to the increased weight.

\begin{figure}
	\includegraphics[width= 1\columnwidth]{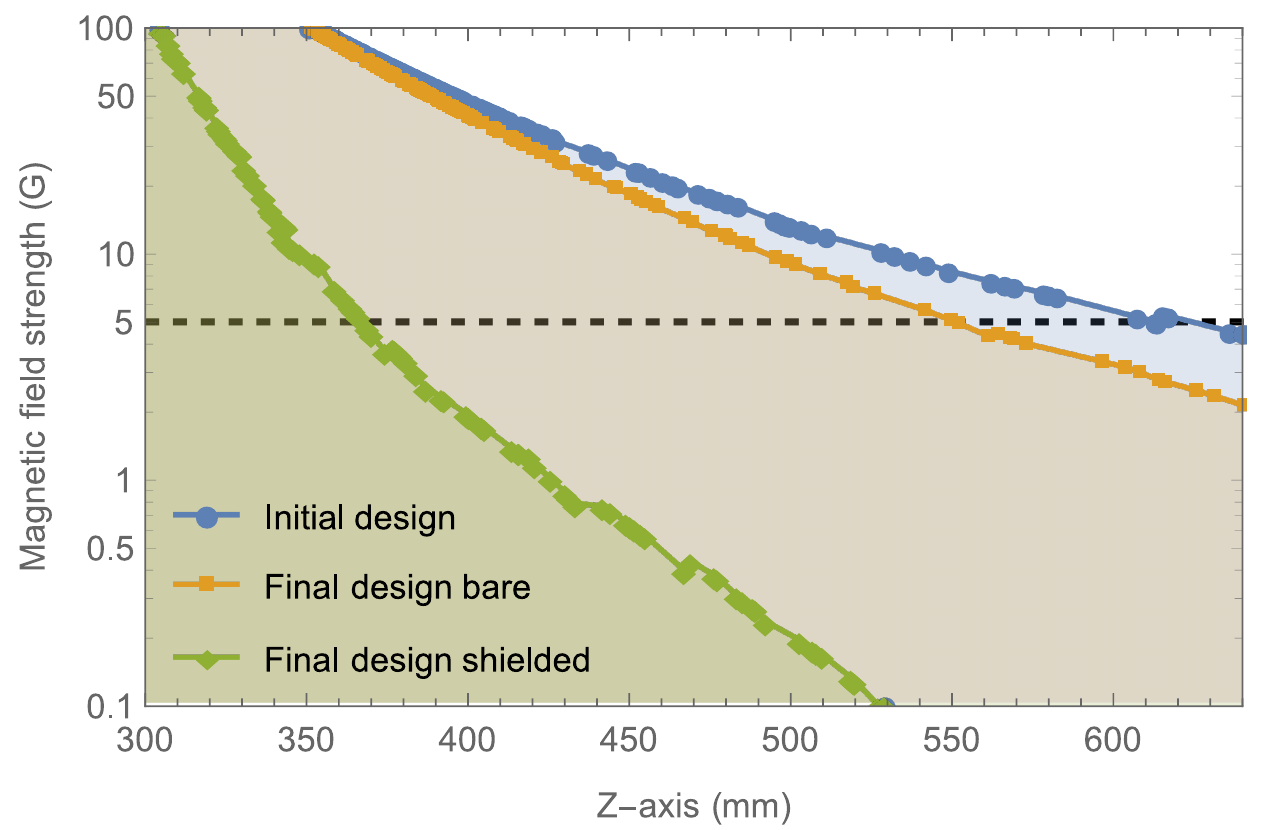}
	\caption{Fringe-field strength simulated along the Z-axis ($Z\geq\SI{300}{mm}$) for a maximal field strength of \SI{1}{T} with and without the magnetic shielding finally installed. The dashed line indicates a field strength of \SI{5}{G}.} 	\label{fig:MagnetFieldStrengths2}
\end{figure}

The solution we finally implemented compresses the magnetic field lines inside the magnet with two steel plates on its front and back. After simulating a variety of arrangements, we opted for steel plates of \SI{10}{mm} thickness separated \SI{20}{mm} from the magnet faces. This results in a \SI{5}{G} surface just beyond the magnet outer walls (Fig.~\ref{fig:FringeDist}, top right). However, the extra weight from the steel plates forced us to redimension the yoke, whose length went from \SI{200}{mm} to \SI{270}{mm} and whose thickness went from \SI{90}{mm} to \SI{67}{mm}. In this final design, the \SI{5}{G} surface is confined to be $<\SI{365}{mm}$ from the magnet center and $<\SI{65}{mm}$ from its outer structure (Fig.~\ref{fig:MagnetFieldStrengths2}). Simulations indicate that the magnetic field homogeneity in the FoV is barely affected by the shield.

\subsection{Cooling system}

\begin{figure}
	\begin{center}
	\includegraphics[width= 0.8\columnwidth]{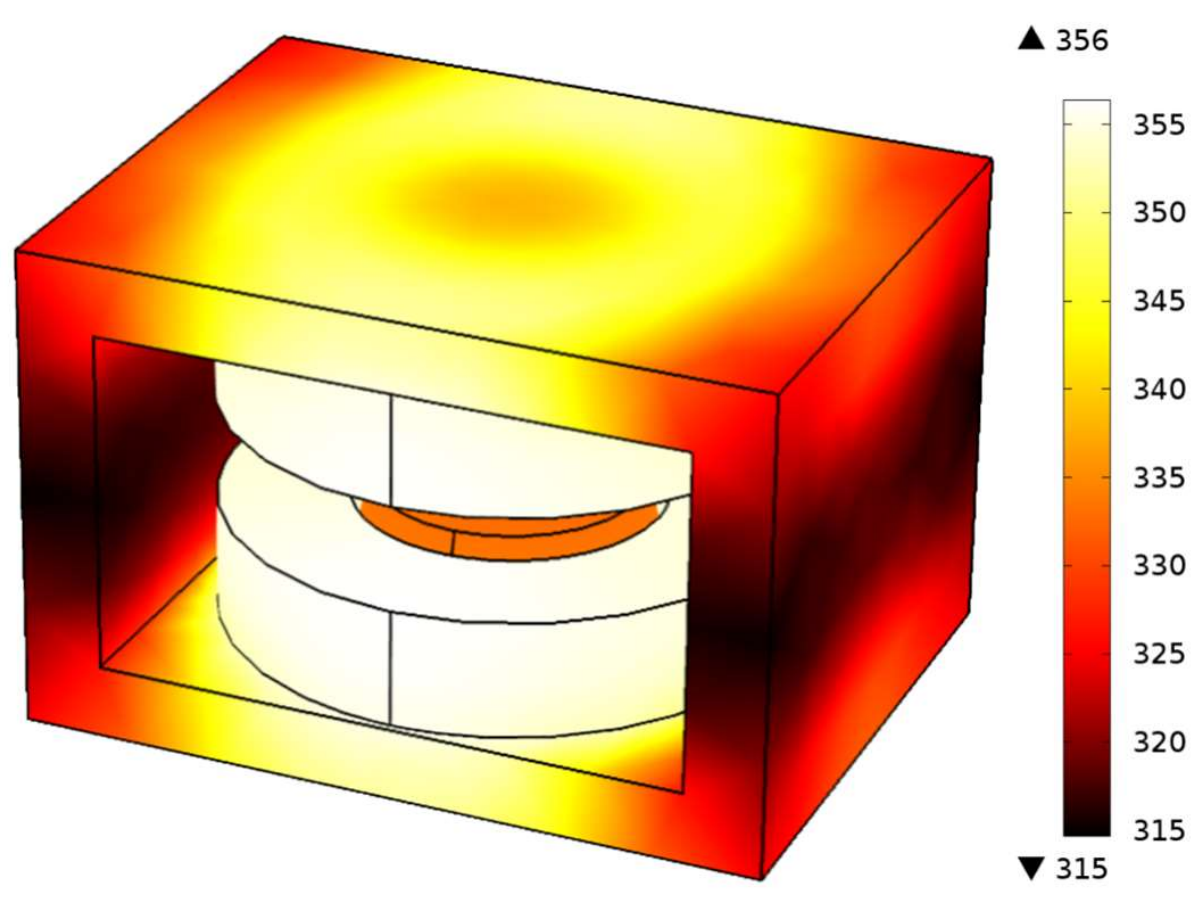}
	\includegraphics[width= 0.95\columnwidth]{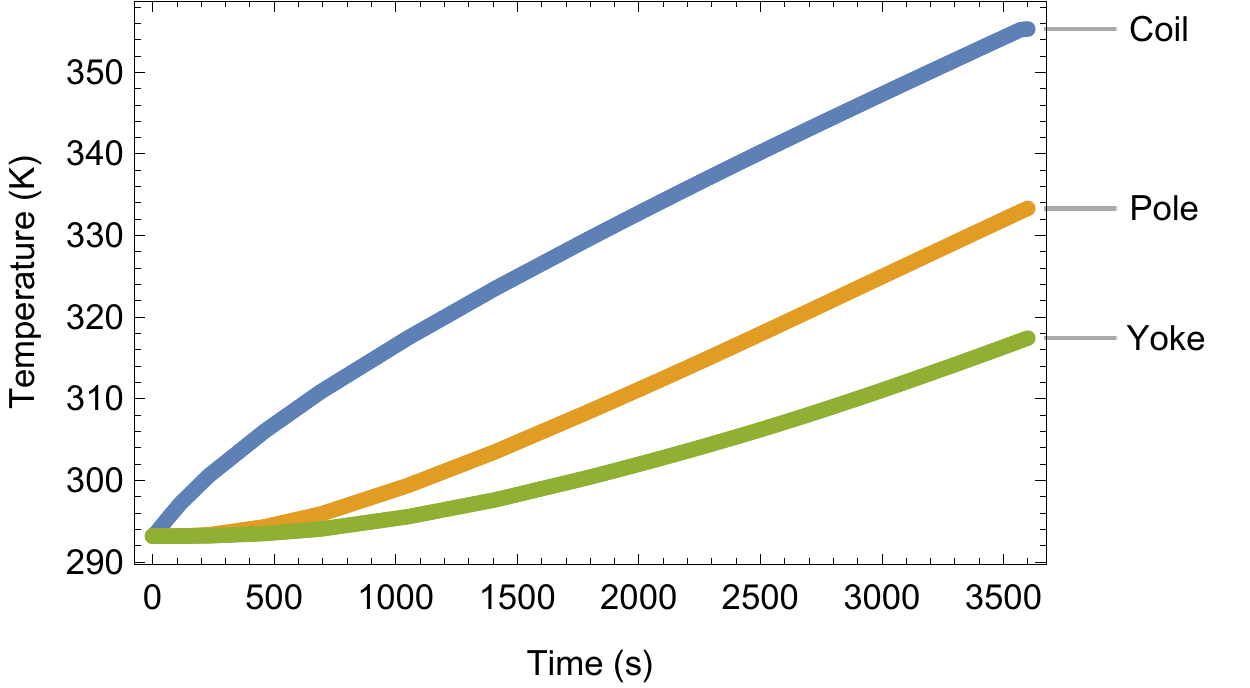}
	\caption{Temperature distribution after (top) and evolution during (bottom) the first hour of operation at $B_0 = \SI{1}{T}$ without cooling. All temperatures are in Kelvin.}
	\label{fig:TempDist}
	\end{center}
\end{figure} 

During operation, the magnet requires currents up to \SI{180}{A}, leading to voltage drops of up to \SI{30}{V} across the coils. Hence, up to \SI{5.4}{kW} of power can be dissipated in the magnet structure. The simulated temperature distribution under these extreme conditions (without cooling) are shown in Fig.~\ref{fig:TempDist}. The resulting equilibrium temperatures are not practical, so we designed a cooling system based on a closed refrigeration loop and a water chiller. 

We have split each coil into four loops, with three layers per loop, to allow for efficient cooling. Thermal simulations indicate that a water flow of \SI{12}{l/min} and a drop pressure of \SI{4}{bar} will limit the temperature increase at the coils to \SI{6}{K}.

\subsection{Shimming System}

The expected homogeneity of \SI{68}{ppm} (Sec.~\ref{sec:magnet}) may prove to be insufficient for high resolution MRI, which could require $<\SI{10}{ppm}$ \cite{12Boer}. In order to meet these demands we require high-order shimming capabilities \cite{12Pan}, which can be delivered by a multichannel system based on an array of planar, fingerprint coils \cite{18Grau}. Further details on the design and performance of the shimming coils and control electronics are given elsewhere \cite{18Puchalt}. 

\subsection{Construction}


The dimensions and weight of the magnet including the shielding system are $594 \times 452 \times \SI{540}{mm^3}$ and \SI{907}{kg}, respectively. The yoke is made of steel St.37 and the poles are XC06. The coil consists of a hollow conductor type with a conductor cross section size $8 \times \SI{8}{mm^2}$ and a circular hole of diameter \SI{4}{mm}. Further details on the magnet assembly are given in the Appendix.

\section{Magnetic field measurements}

\begin{figure*}
	\centering
	\includegraphics[width= 1.5\columnwidth]{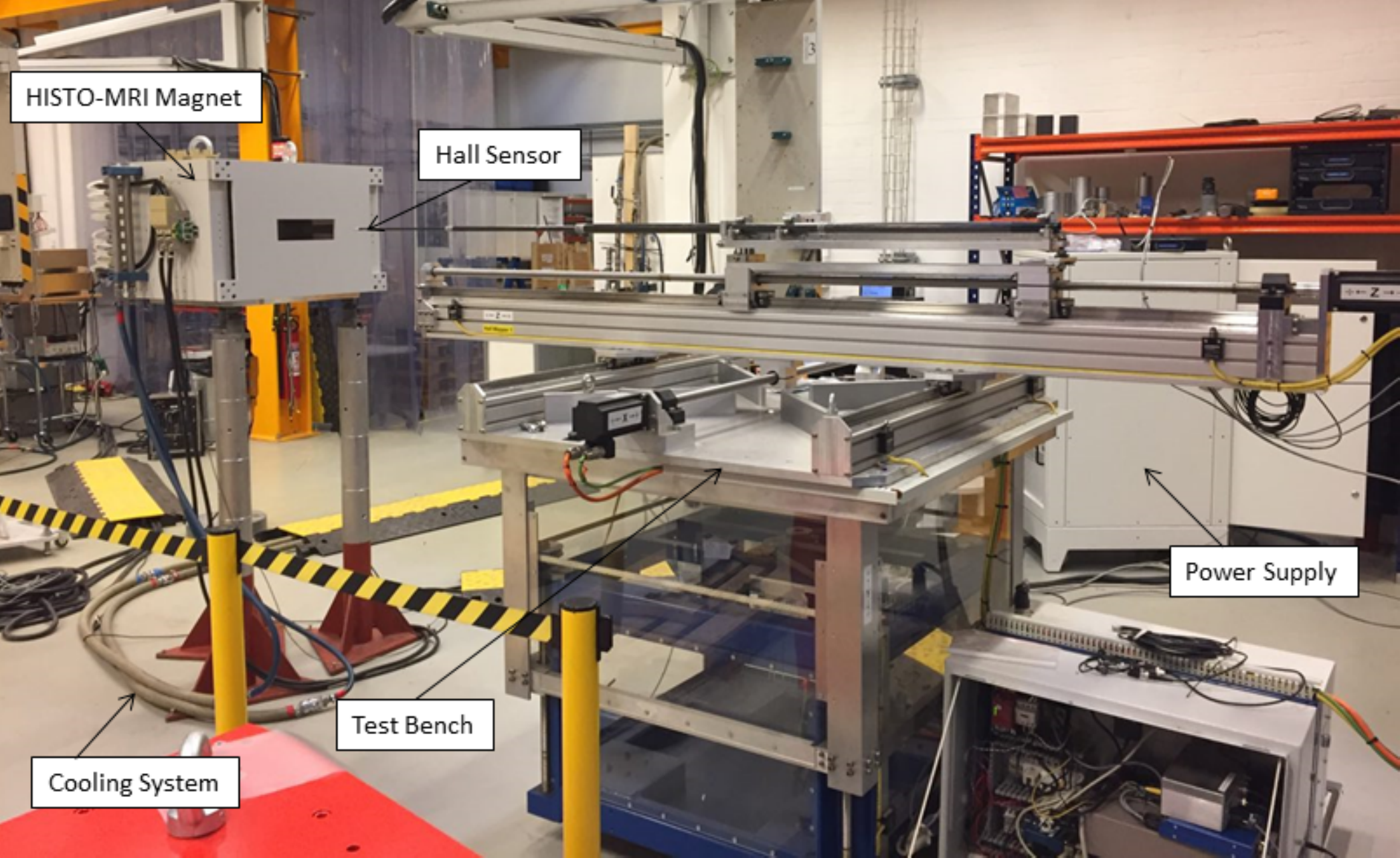}
	\caption{HISTO-MRI magnet and magnetic test bench at Danfysik S/A factory.}
	\label{fig:HallTestBench}
\end{figure*} 

We have characterized the spatial field-strength distribution and overall performance of the main magnet with a Hall probe attached to a 3D positioning system. The positioners in the transverse (vertical) directions move in a range of up to \SI{1.3}{m} (\SI{0.36}{m}) and are accurate at the \SI{100}{\micro m} level, whereas the Hall probe is specified to have an accuracy of \SI{30}{mT} with the settings used in most measurements. For these tests we supplied current to the magnet with a Danfysik 9100 Unipolar Power Supply (see Fig.~\ref{fig:HallTestBench}).

\begin{figure}
	\centering
	\includegraphics[width= 0.9\columnwidth]{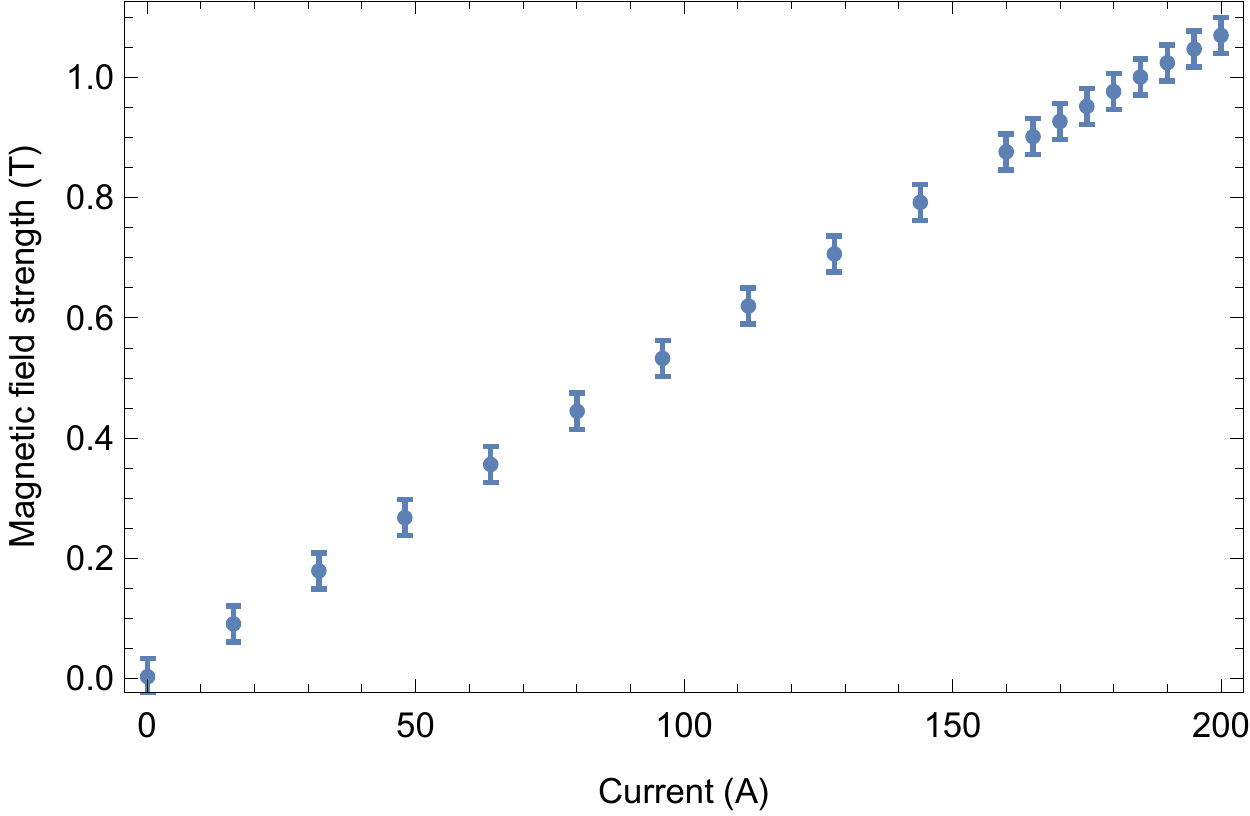}
	\caption{Magnetic field strength at the center of the FoV measured as a function of the current.}
	\label{fig:LinearityError}
\end{figure}

An important specification of an electromagnet is the linearity of the generated field with respect to the current run through its coils. Non-linearities can appear due to mechanical stress in the coils from thermal effects, as well as loss in the iron poles. To characterize these effects, we placed the Hall probe at the center of the FoV and ramped the current up to the nominal maximum. The measured results (Fig.~\ref{fig:LinearityError}) show that $B_0\approx\SI{1}{T}$ for a current $\approx\SI{186}{A}$ (and a voltage of $\approx\SI{30}{V}$ across the coil), close to the design value. Measurements of the field at the center of the FoV are shown in Fig.~\ref{fig:LinearityError} as a function of the input current. At \SI{1}{T} the non-linearity is $\approx2~\%$ due to a modest iron-loss. Magnetic hysteresis (iron memory effect) is $\approx0.4~\%$ and can be avoided with current conditioning and always ramping the current up from zero to the set point.

\begin{figure}
	\centering
	\includegraphics[width= 0.95\columnwidth]{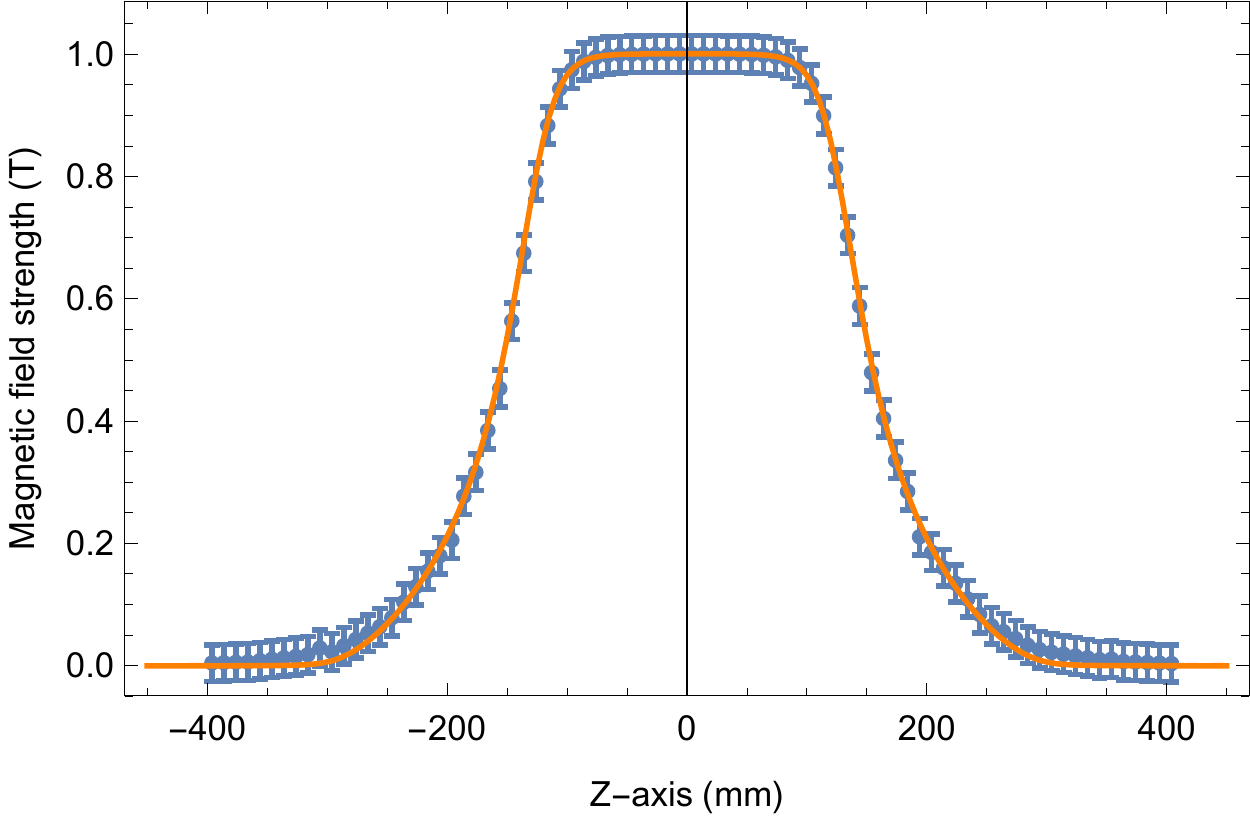}
	\includegraphics[width= 0.95\columnwidth]{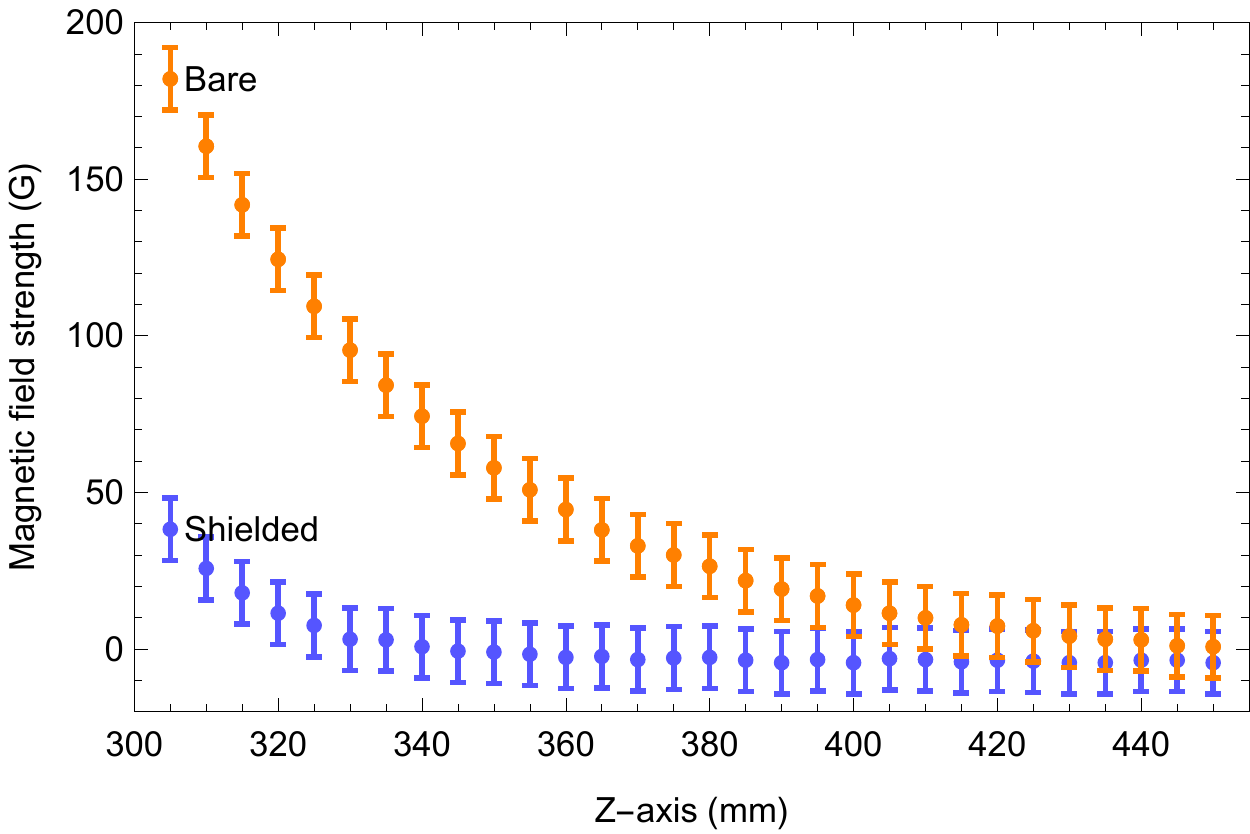}
	\caption{(Top) Magnetic Field profile along the Z-axis for $B_0\approx\SI{1}{T}$ for COMSOL simulations (solid, orange line) and experimental measurements (blue points). (Bottom) Fringe-field measurements along the Z-axis for a maximal field strength of \SI{1}{T} for the final magnet design with and without the shielding. The nominal uncertainty of the Hall sensor in the scale used for these measurements is \SI{10}{G}.}
	\label{fig:COMSOLExperimental}
\end{figure}

The magnetic field strength measured along the Z-axis closely resembles that expected from electromagnetic simulations (Fig.~\ref{fig:COMSOLExperimental}). We have also measured the field with and without the magnetic shield to characterize its influence on the position of the \SI{5}{G} line and the homogeneity in the FoV (Fig.~\ref{fig:COMSOLExperimental} bottom). These measurements indicate that the 5~G fringe-field limit for the magnet without the shield lies $\approx\SI{430}{mm}$ from the magnet center ($\approx\SI{130}{mm}$ from the outer surface), while placing the shield brings it down to $\approx\SI{325}{mm}$, in agreement with simulations. The effect of the shield on the field distribution in the FoV was imperceptible.

\begin{figure}
	\centering
	\includegraphics[width= 1.\columnwidth]{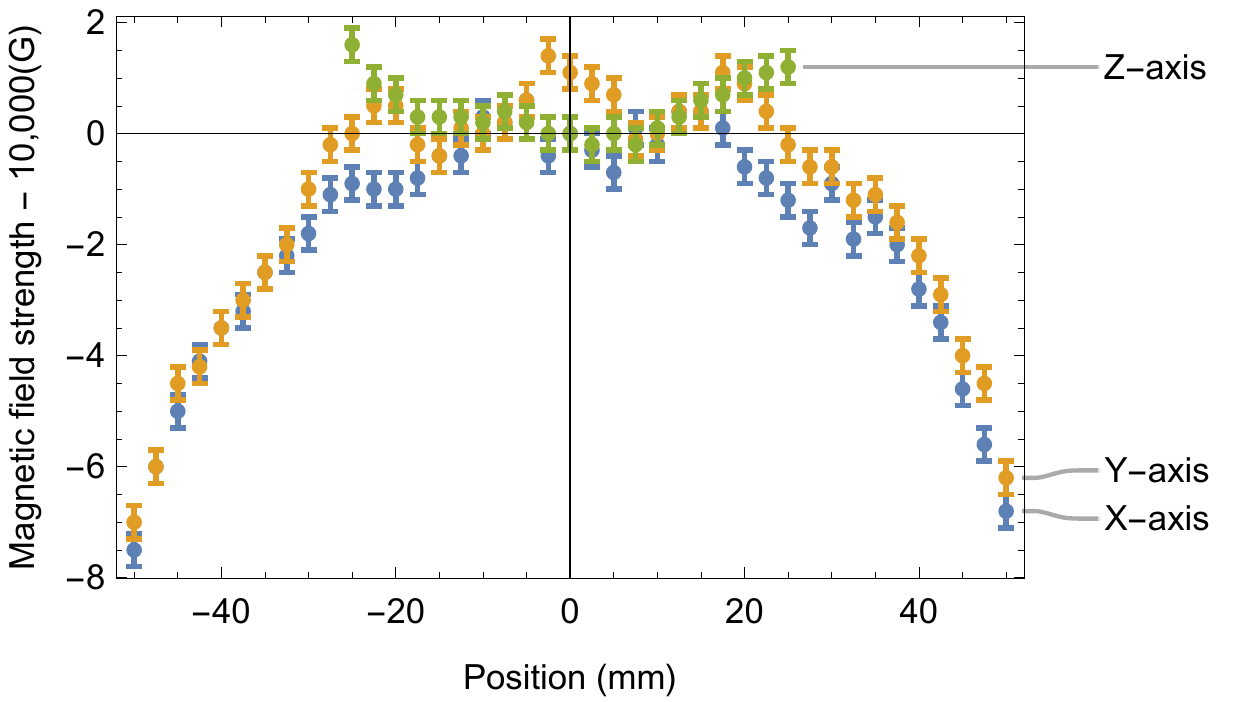}
	\caption{Magnetic field strength along the X (blue), Y (green) and Z (red) axes for $B_0\approx\SI{1}{T}$. }
	\label{fig:Homogeneity}
\end{figure}

Regarding homogeneity, the magnet was designed for relative $B$-field variations $<\SI{100}{ppm}$ in a FoV of radius \SI{10}{mm}. We scanned a cubic volume of sides \SI{20}{mm} and measured an homogeneity $\approx\SI{71}{ppm}$. The corresponding simulated value is $\approx\SI{68}{ppm}$. Deviations from the nominal $B_0$ along the three Cartesian axes can be read from Fig.~\ref{fig:Homogeneity}.

\section{Conclusion}
We have designed, built and characterized a compact electromagnet (\SI{<1000}{kg}) capable of generating a variable magnetic field between 0.5 and \SI{1}{T}, while keeping the magnetic field inhomogeneity \SI{<100}{ppm} in a spherical FoV of \SI{20}{mm} diameter. This magnet is a crucial component of a unique MRI scanner which aims at demonstrating a technology capable of in-vivo images of deep tissues with histological spatial resolution \cite{HISTOMRI}.

\appendix[Mechanical structure]
\label{sec:mechanical}

\begin{figure}
	\centering
	\includegraphics[width= 0.75\columnwidth]{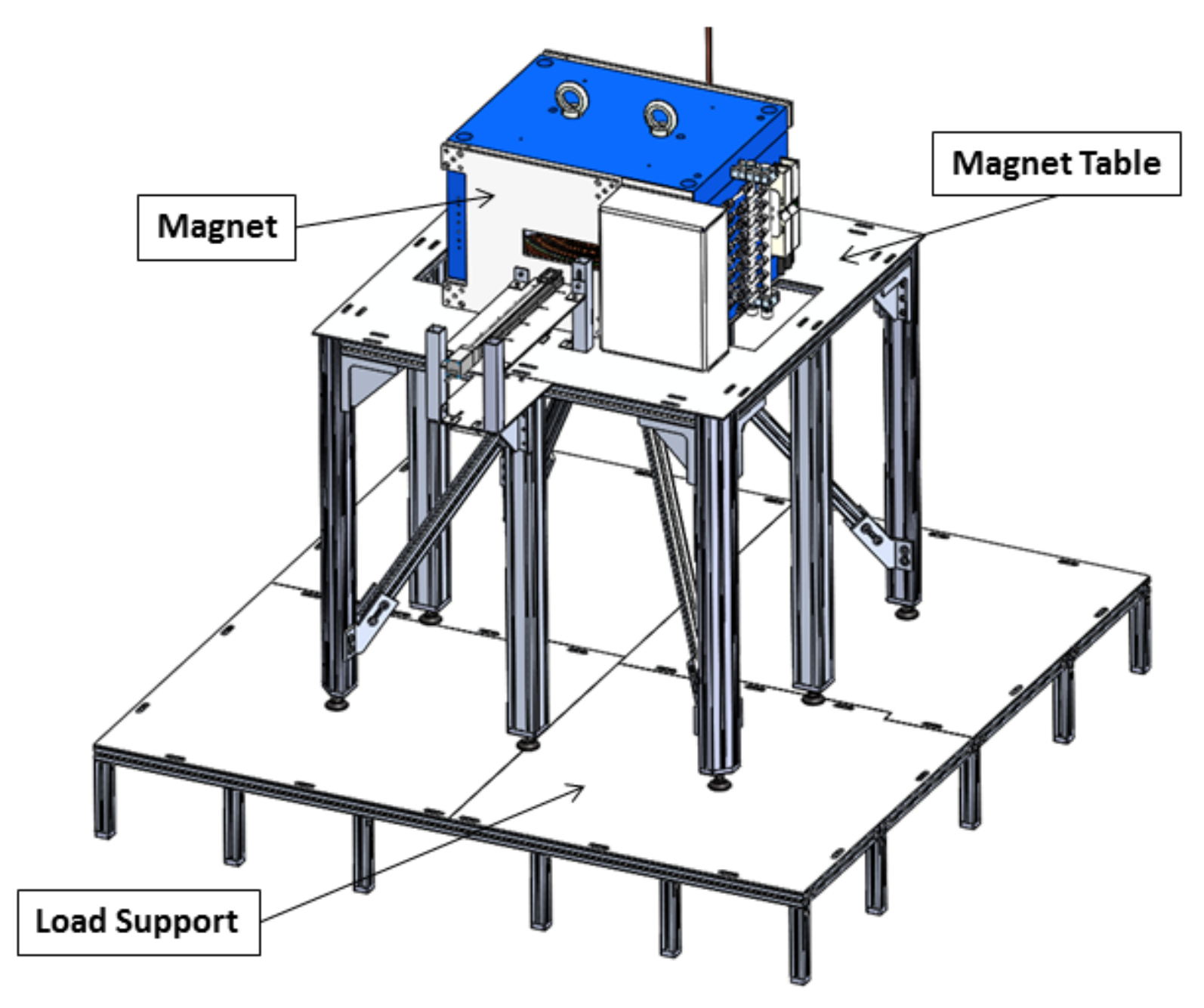}
	\caption{3D CAD view of the magnet and the mechanical support.}
	\label{fig:3DCADMechanicalModel}
\end{figure}

The weight of the magnet imposes the use of a robust and reliable mechanical structure. To ensure the structural integrity of the system and the building, the load must be kept <\SI{350}{kg/m^{2}} and the load distribution should be as uniform as possible.

To avoid the strong forces between the support structure and the magnet, we built the former out of aluminum 6063. The mechanical structure of the magnet is formed by two components: the magnet table and the load support (Fig.~\ref{fig:3DCADMechanicalModel}). The table holds the electromagnet and other components of the MRI scanner, including the platform for moving the sample in and out of the FoV. It spreads the reaction force of the electromagnet uniformly over a plate of \SI{1}{m^2} via 8 legs, 4 cross-bars and 9 framing squares. This structure rests on the lower load support, which has a surface of \SI{4}{m^2} and consists of 24 legs and 4 aluminum plates. Nine beams across the aluminum plates limit material deformations.

\begin{figure}
	\centering
	\includegraphics[width= 1.\columnwidth]{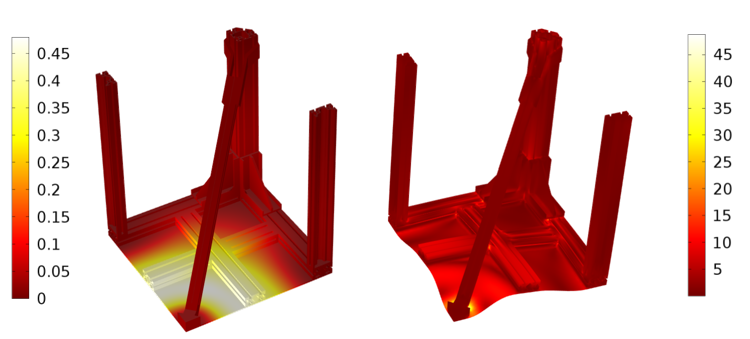}
	\caption{(Right) Total deformation (mm) and (Left) Von Mises stresses for the magnet table (MPa).}
	\label{fig:COMSOLMagnetTable}
\end{figure}

\begin{figure}
	\centering
	\includegraphics[width= 1.\columnwidth]{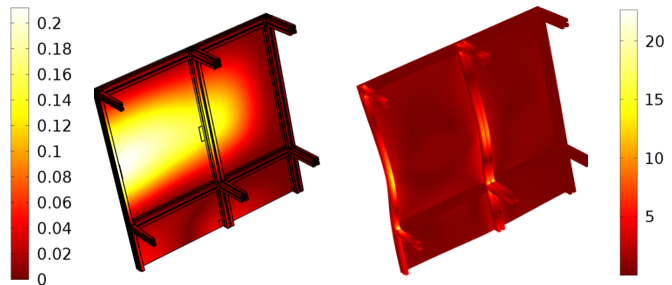}
	\caption{(Right) Total deformation (mm) and (Left) Von Mises stresses for the load support (MPa).}
	\label{fig:COMSOLLoadSupport}
\end{figure}

\begin{figure}
	\centering
	\includegraphics[width= 0.9\columnwidth]{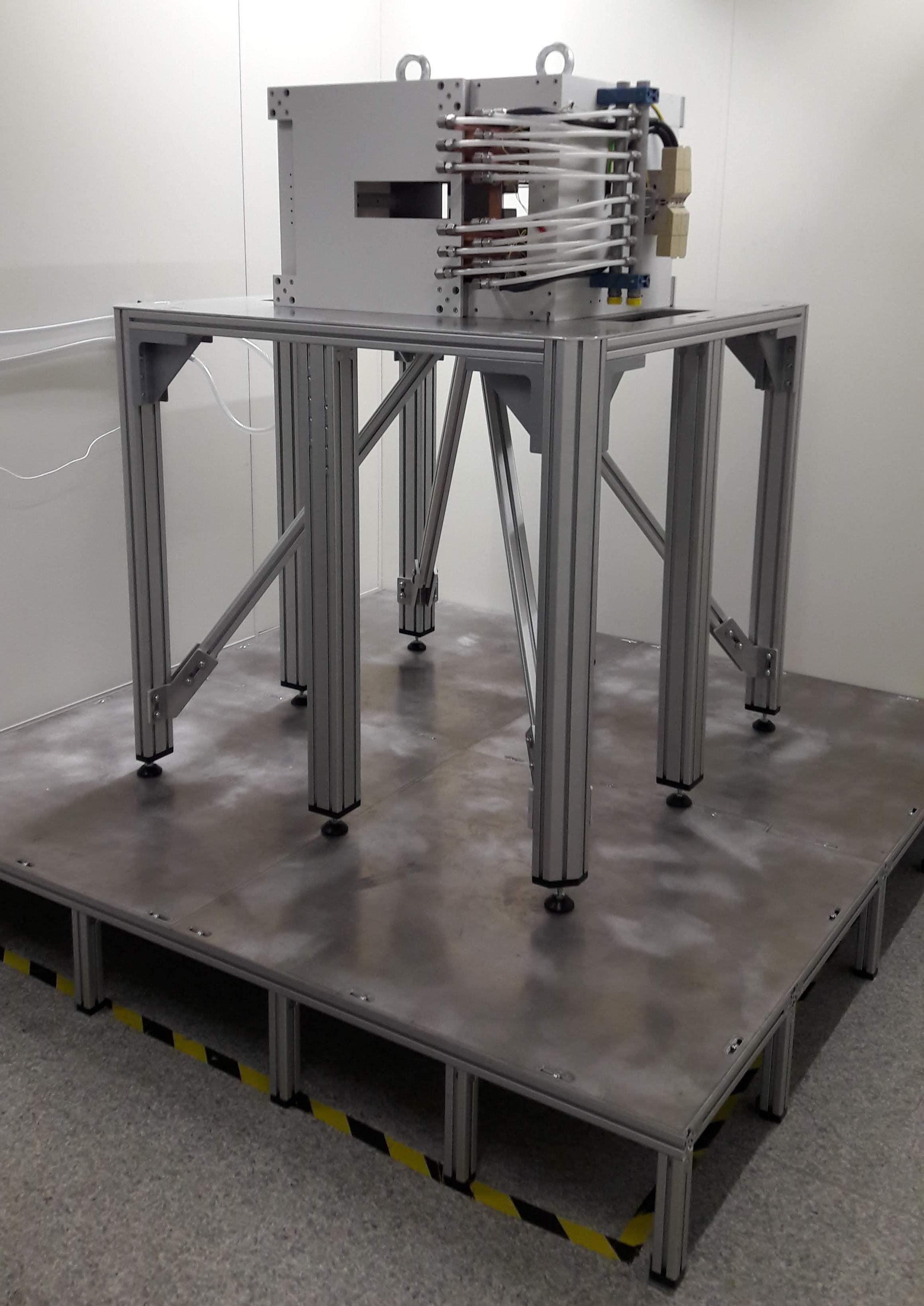}
	\caption{Picture of the Magnet installed in the MRILab (Valencia, Spain).}
	\label{fig:MagnetLAB}
\end{figure}

We have analyzed the structure mechanically to characterize deformations, stresses and reactions. To simplify and speed up the analysis we simulate half of the structure and profit from the present symmetries. The load due to the magnet ($594 \times 452 \times \SI{540}{mm^3}$) amounts to $\approx\SI{10}{kN}$. The reaction force on the table legs due to gravity is $\approx\SI{1.35}{kN}$ per leg. The results of the Von Mises Stress and the maximum deformation for the magnet table are shown in Fig.~\ref{fig:COMSOLMagnetTable}. The total weight is $\approx\SI{1.25}{T}$ spread over $\SI{4}{m^{2}}$ on the load support structure, thereby complying with building specifications. The assessment shows that the yield limit of the structure, that has $\SI{46}{MPa}$ as maximum value, will not exceed the maximum permissible value.  The mechanical structure has a safety factor of 4 and shall withstand the weight. The results of the Von Mises Stress and the maximum deformation for the load support are shown in Fig.~\ref{fig:COMSOLLoadSupport}. A photograph of the mechanical support mounted in the MRILab at the i3M is shown in Fig.~\ref{fig:MagnetLAB}.

%

\section*{Acknowledgment}

We thank the Infrastructures Department of Universidad Politécnica de Valencia for their help regarding storage, installation and assembly of the magnet. This work was supported by the European Commission under Grant 737180 (HISTO-MRI).

\ifCLASSOPTIONcaptionsoff
  \newpage
\fi

\bibliography{myrefs}

%




\end{document}